\documentclass{aa}
\usepackage{graphics,epsf,epsfig,graphicx}

\begin{document}

\title{Gravitational torques in spiral galaxies:\\
gas accretion as a driving mechanism of galactic evolution}
                                
\author{David L. Block \inst{1}, Fr\'ed\'eric Bournaud \inst{2,3},
Fran\c coise Combes \inst{2}, Iv\^anio Puerari \inst{4}, Ronald Buta \inst{5}}
 
\offprints{D. L. Block \email{block@cam.wits.ac.za}} 
\institute{
       School of Computational and Applied Mathematics,
       University Witwatersrand, Private Bag 3, WITS 2050, South Africa
\and
      Observatoire de Paris, LERMA, 61 Av. de l'Observatoire, F-75014, Paris, 
France
\and
      Ecole Normale Sup\'erieure, 45 rue d'Ulm, F-75005, Paris, France
\and
      Instituto Nacional de Astrof\'\i sica, Optica y Electr\'onica,
      Calle Luis Enrique Erro 1, 72840 Tonantzintla, Puebla, M\'exico
\and
      Dept. of Physics and Astronomy, University of
      Alabama,  Box 870324, Tuscaloosa, Alabama 35487, USA } 
\date{Received XX XX, 2002; accepted XX XX, 2002}
\authorrunning{Block et al.} 
\titlerunning{Gravitational torques, gas accretion and galaxy evolution}

\abstract{The distribution of gravitational torques and bar strengths
in the local Universe is derived from a detailed study 
of 163 galaxies observed in the near-infrared. The results are compared 
with numerical models for spiral galaxy evolution. It is found that the 
observed distribution of torques can be accounted for only with external 
accretion of gas onto spiral disks. Accretion is responsible for bar renewal 
-- after the dissolution of primordial bars -- as well as the maintenance of 
spiral structures. Models of isolated, non-accreting 
galaxies are ruled out. Moderate accretion rates do not explain the 
observational results: it is shown that galactic disks should double their 
mass in less than the Hubble time. The best fit is obtained if spiral 
galaxies are open systems, still forming today by continuous gas accretion, 
doubling their mass every 10 billion years.
}

\maketitle
\section{Introduction}
Bars are a major perturbation of
the gravitational potential and a highly efficient engine for the evolution of
morphological and chemical properties. In purely stellar disks, bars
are robust, 
long-lived structures (Combes \& Sanders 1981). In reality, however,
spiral galaxies contain gas 
which provokes the dissolution of bars (Bournaud \& Combes 2002, hereafter BC). 
The bar itself initiates an
important radial gas inflow, which destroys 
barred structure (Pfenniger \& Norman 1990). 

Since bars in gaseous disks are dissolved in a few Gyrs, the ubiquity of bars 
in the local Universe suggests that bars are also renewed. 
BC painted a scenario wherein continual 
gas accretion might be a crucial evolutionary factor. 
This gas settles in the disk, enhancing its self gravity, 
reducing the influence of the stabilizing central mass 
concentration, and a second bar develops. In this Letter, we test 
this hypothesis, by comparing the observed distribution of gravitational 
torques (in a well-defined galaxy sample) with the distribution predicted by 
numerical simulations incorporating gas accretion.

\section{Observational and numerical techniques}\label{tech}
\subsection{Gravitational torques}
The maximal torque over the entire disk is defined by
\begin{equation}
Q_\mathrm{T}(R) = {F_\mathrm{T}^{\mathrm{max}}(R) \over F_\mathrm{0}(R)} = 
{{{1\over R}\bigl{(}{\partial \Phi(R,\theta)\over \partial\theta}\bigr{)}_
{\mathrm{max}}} \over {d\Phi_\mathrm{0}(R)\over dR}} 
\end{equation}
\noindent
where $F_\mathrm{T}^{\mathrm{max}}(R)$ 
represents the maximum amplitude of the tangential force
and $F_\mathrm{0}(R)$ is the mean axisymmetric radial force, inferred
from the axisymmetric component of the gravitational potential, $\Phi_\mathrm{0}$. The potential
is inferred from the visible mass only, in both simulations and observations. 
We then measure $Q_\mathrm{b}=\max_{R}(Q_\mathrm{T}(R))$. $Q_\mathrm{b}$ is simply the bar strength in 
barred galaxies, or the maximal arms torque in unbarred or nearly unbarred 
galaxies.

\subsection{Observations}
To derive the distribution of $Q_\mathrm{b}$, we use near-infrared images from the 
Ohio State University Bright Galaxy Survey (OSUBGS, Eskridge et al. 
2002). The full sample consists of 198 spiral galaxies having total 
apparent blue magnitude $B_\mathrm{T}$ $<$ 12 and blue isophotal diameter 
$D_{25}$ $<$ 6$^{\prime}$. The technique for deriving $Q_\mathrm{b}$ from near-IR 
images is described by Buta \& Block (2001; see also Block et al. 2001), 
who used the method of Quillen, Frogel \& Gonz\'alez (1994) to transform 
deprojected, cleaned images into potentials.

Of the 198 OSUBGS galaxies, only 163 are used here for studies of 
gravitational torques, for reasons detailed in Sect.~\ref{41}. A constant 
mass-to-light ratio is assumed, i.e, the dark halo is ignored. The disk 
scale-height and scale-length are assumed to be related by $h_\mathrm{z}=h_\mathrm{r}/12$. 
Also, bulges are assumed to be as flat as the disk; no decompositions to 
account for the likely rounder shapes of some bulges have been carried 
out for this study.
Such refinements will be considered in a separate study (Buta et al.
2002).

\subsection{Numerical simulations}
We employ a numerical code to study the evolution of bars and spiral arms 
in galaxies which do, or do not, accrete gas. 
The simulations include gas dynamics, star formation, and stellar
mass-loss. They model a disk, a bulge, and a dark halo. 
For a complete description, see BC. 

We simulate the evolution of four model galaxies,  assumed to represent 
types Sa, Sb, Sc and Sd, with parameters given in Table~\ref{params}. 
The proportion of each morphological type in the observed sample is fully 
accounted for when we analyze the results. We compute the evolution of 
each model over 15 Gyr. We calculate the mean value of $Q_\mathrm{b}$ at periods of 100 Myr.

\begin{table}
\centering
\begin{tabular}{cccc}
\hline
\hline
Type & bulge-to-disk & halo-to-disk & abundance   \\
     &   mass ratio  &  mass ratio  & (OSU sample) \\
\hline
 Sa  &     0.9       &     0.3      &      17\%     \\
 Sb  &     0.4       &     0.5      &      25\%     \\
 Sc  &     0.2       &     0.7      &      45\%     \\
 Sd  &     0.1       &     0.9      &      13\%     \\
\hline
\end{tabular}
\caption{Bulge, disk, and halo parameters used in numerical simulations. The 
halo mass that is accounted for is the mass of dark matter inside the disk 
radius.}
\label{params}
\end{table}

\section{Results}\label{res}
\subsection{Distribution of gravitational torques}
\begin{figure}
\centering
      \includegraphics[angle=270,width=7.5cm]{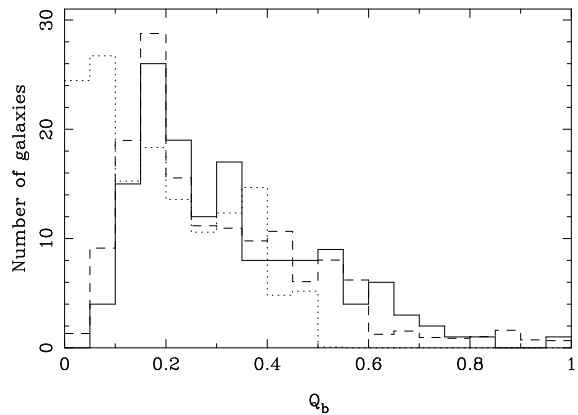}
      \caption{Histogram of gravitational torques. Solid: observations 
      -- Dashed: simulations with gas accretion that double the disk mass in 10 Gyrs 
      -- Dotted: simulations without gas accretion.}
      \label{histo1}
\end{figure}

A histogram showing the number of galaxies as a function of their 
maximal torque is presented in Fig.~\ref{histo1}. 
The most striking feature of the histogram is the depression 
in the number of galaxies with very low torques ($Q_\mathrm{b}\sim 0$). 
That the drop at $Q_\mathrm{b}\sim 0$ is not due to selection effects, but
is indeed robust, is discussed in Sect.~\ref{41}.

In Fig.~\ref{histo1}, one clearly sees that the shape 
of the observed histogram is not reproduced without gas accretion. 
Only large accretion rates 
reproduce the observed histogram (both the drop at $Q_\mathrm{b}\sim 0$ and 
the number of strong bars with $Q_\mathrm{b}>0.5$), when the disk 
mass doubles in 13, 10 or 8 Gyrs. 

We also compute the deviation between the observed and simulated 
histograms. If $N_i$ is the number of galaxies observed in each 
torque class $i$,and $N'_i$ the corresponding value from 
simulations,we define $\sigma^2=\sum_{i} (N_i-N'_i)^2$. 
We normalize $\sigma$ to 1 for non accreting models.
Values of $\sigma$ are listed in Table~\ref{tabsigma}.
The best fit to the observed histogram is obtained when galaxies 
double their mass every 10 Gyrs, while low accretion rates in which 
the disk mass doubles in 30 Gyrs, are ruled out. 

\begin{table}
\centering
\begin{tabular}{cccccc}
\hline
\hline
Accretion rate &  $Q_\mathrm{b}$  & $Q_\mathrm{b}^m$ &  $Q_2$  & $Q_2^m$ & $F_{\mathrm{bar}}$ \\
\hline
 no accretion  &  1.00   & 1.00    &  1.00   &  1.00   &  1.00     \\
 $M_0/30$      &  0.90   & 0.85    &  0.86   &  0.92   &  0.89     \\
 $M_0/13$      &  0.37   & 0.45    &  0.51   &  0.39   &  0.32     \\
 $M_0/10$      &  0.34   & 0.44    &  0.48   &  0.34   &  0.26     \\
 $M_0/8 $      &  0.40   & 0.88    &  0.52   &  0.36   &  0.28     \\
\hline
\end{tabular}
\caption{Values of $\sigma$ for different accretion rates (in Gyr$^{-1}$), and
several gravitational torques and bar strength indicators. $M_0$ is the initial
disk mass.}
\label{tabsigma}
\end{table}

\subsection{Interpretation: temporal bar evolution}
The presence of gas in galactic disks is responsible for the destruction
of bars in no more than 5 Gyrs. As elucidated by BC, gas is also responsible for bar renewal, when it is accreted from
outside the disk. 
Gas accretion radically changes the temporal evolution of the bar
strength: see for example Fig. 4 in BC. 
With accretion, once a bar is dissolved, the disk can become unstable 
again and a new bar may form. 
The disk rarely resides in axisymmetric states: 
even between bar episodes, spiral arms maintain a significant 
torque, for accretion also rejuvenates the spiral structure.
Without accretion, the disks spend half of their lifetimes in
nearly axisymmetric states with $Q_\mathrm{b}<0.05$: both bars and
arms disappear. The number of galaxies in
each class of the histogram in Fig.~\ref{histo1} is
interpreted as the time fraction galaxies spend in each class during
their lives.

\section{Discussion}\label{disc}
\subsection{Determination of gravitational torques}\label{41}
In deriving gravitational torques, we assume that each disk 
has a constant scale-height.
What we measure is thus different from the actual torque in nature. 
However, we compute $Q_\mathrm{b}$ exactly in the same manner 
for the simulations. We adopt the same set of assumptions, 
which allows a fair comparison between observed and simulated distributions 
to be made. Even if the value of $Q_\mathrm{b}$ is different from the actual maximal torque, the same error will follow through in our simulations. 

Obviously, $Q_\mathrm{b}$ will differ from the actual torque also because of the dark matter potential. However, the maximal torque is obtained well inside the disk, where the dark matter contribution is negligible. Moreover, $Q_\mathrm{b}$ is also computed in simulations ignoring dark matter.

A differentiation between bar and spiral arm torques is not needed.
Spiral structure is also present in simulations, so knowing whether 
$Q_\mathrm{b}$ is related to a bar or to a spiral structure is not required.

As mentioned above, only 163 galaxies from 198 in the OSU sample 
were retained. Our selection criteria are delineated as follows:
\begin{itemize}
\item galaxies inclined more than 70 degrees  cannot be 
properly deprojected (deprojection is discussed later)
\item members of obvious interacting systems are not studied
\end{itemize}
It is unlikely that these criteria are responsible for the depression 
in the number of galaxies with low torques: in particular, unbarred 
galaxies were not given preferential rejection.

Thus far, we have itemized several approximations in $Q_\mathrm{b}$, 
but the same assumptions have followed through in the simulations, 
so that the comparison remains fair. The most 
questionable point is in fact deprojection, for only observations 
require such a treatment. An axisymmetric disk 
could appear non-axisymmetric when the deprojection is not 
correct: here some approximations may be introduced in 
observations, but not in simulations. 
First, observational deprojection assumes that outer disks are circular, which simulations suggest is unlikely. To estimate the resulting uncertainty, we experimented by randomly orienting the simulated images, and deprojecting them exactly as 
was done for observations. If we denote by $\Delta Q_\mathrm{b}$ the change introduced 
on each bar torque value, we find from simulations 
that $\sqrt{<{\Delta Q_\mathrm{b}}^2>}=.022$, 
and $<\Delta Q_\mathrm{b} >=.007$. A second problem is 
related to the observational uncertainty of the disk axis ratio $b/a$. We use 
data from LEDA to determine the uncertainty on $b/a$ as a function of $b/a$. 
We then convert the result into an uncertainty upon $Q_\mathrm{b}$ as a function of the 
estimated inclination $i$ of the disk, and add to this the first uncertainty 
on $Q_\mathrm{b}$, 
related to the outer disk shape; values are given in Table~\ref{tabQB}. 
Given that the bin of our histogram is 0.05, the inclination cutoff at 70 
degrees is justified, and the results of deprojection are dependable.
Another deprojection issue is related to the bar orientation relative to 
the line of sight: this may affect strongly barred galaxies, but 
will not affect the crucial number of galaxies with low torques.

\begin{table}
\centering
\begin{tabular}{cccccc}
\hline
\hline
$i$ &$\Delta Q_\mathrm{b}$& $i$ &$\Delta Q_\mathrm{b}$& $i$ &$\Delta Q_\mathrm{b}$  \\
\hline
 20  &  0.022    & 50  &  0.025     &  70   &   0.039    \\
 35  &  0.023    & 60  &  0.028     &  80   &   0.052    \\
\hline
\end{tabular}
\caption{Deprojection uncertainties on gravitational torques $\Delta Q_\mathrm{b}$, as a function of the disk inclination $i$.}\label{tabQB}
\end{table}

\subsection{Other measures of torques and bar strengths}

It is interesting to examine results obtained with physical indicators other than
$Q_\mathrm{b}$. As mentioned above, $Q_\mathrm{b}$ is related either to a bar or to arms, for we select the maximal torque.
It may seem better to account for both bar and arms in every disk, by 
measuring the mean torque over the whole disk:
\begin{equation}
Q_\mathrm{b}^\mathrm{m}=\frac{\int_{r_\mathrm{min}}^{r_\mathrm{max}}  Q_\mathrm{T}(r) r dr}{\int_{r_\mathrm{min}}^{r_\mathrm{max}} r dr}
\end{equation}
where $r_\mathrm{max}$ is the optical radius and $r_\mathrm{min}$ is 7\% of $r_\mathrm{max}$ 
(to avoid central singularities). 
One could also try to eliminate spiral structures and to preferentially focus on bars. To accomplish this, instead of measuring the maximal 
torque, we measure the torque in the $m$=2 Fourier component: 
if the potential is decomposed as
\begin{equation}
\Phi(r,\theta) = \Phi_0(r) + \sum_m \Phi_m(r) \cos (m \theta - \phi_m)
\end{equation}
we define $Q_2(r) = 2 \Phi_2 / r | F_0(r) |$ and $Q_2=\max_r Q_2(r)$. 
This eliminates spirals with more than 2 arms, thus this parameter may 
usually be regarded as a measure of bar strength (Combes \& Sanders 1981). 
Finally, one may incorporate the spatial extent of bars by measuring:
\begin{equation}
Q_2^\mathrm{m}=\frac{\int_{\mathrm{r_\mathrm{min}}}^{\mathrm{r_\mathrm{max}}}  Q_2(r) r dr}{\int_{r_\mathrm{min}}^{r_\mathrm{max}} r dr}
\end{equation}

In contrast, Whyte et al. (2002), have determined the distribution 
of bar strengths from the OSU sample using bar shapes. They define their 
bar strength parameter by:
\begin{equation}
F_{\mathrm{bar}}=\frac{2}{\pi}\left[\arctan(b/a)^{-\frac{1}{2}} - \arctan(b/a)^{\frac{1}{2}} \right]
\end{equation}
where $b/a$ is the axis ratio of the bar, 
i.e. the minimum value of the axis ratio of isophotes in an ellipse-fitting 
model (Abraham et al. 1999). Whyte et al. (2002) also find that nearly 
axisymmetric disks are rare.

\begin{figure}
\centering
      \includegraphics[angle=270,width=7.5cm]{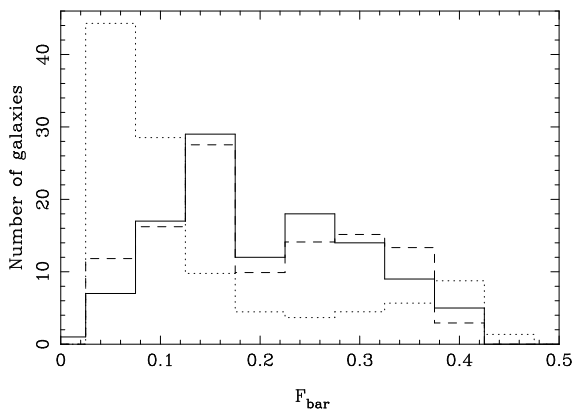}
      \caption{Histogram of bar strength $F_{\mathrm{bar}}$. Solid: observations, data from Whyte et al. (2002)  
      -- Dashed: simulations with gas accretion that doubles the disk mass in 10 Gyrs 
      -- Dotted: simulations without gas accretion.}
      \label{histo2}
\end{figure}

It is difficult to justify  one indicator instead of another 
one; indeed it is important to test our accretion hypothesis 
with several indicators. 
The results are given in Table~\ref{tabsigma}. We also show 
histograms of $F_{\mathrm{bar}}$ in Fig.~\ref{histo2}. For other indicators ($Q_2$, $Q_2^\mathrm{m}$, $Q_\mathrm{b}^\mathrm{m}$), histograms still show a drop at $Q \sim 0$, which is only reproduced by models with accretion.

All these additional indicators confirm 
the previous results: the properties of spiral galaxies cannot be 
reproduced without gas accretion, but require galaxies to double their mass 
in less than a Hubble time. The best accretion rate makes the mass 
of spiral disks double over epochs spanning about 10 Gyrs. The difference 
between the observations and models of closed systems -- or galaxies 
which accrete gas only at a marginal rate -- is so flagrant that 
such models are clearly ruled out. Only large accretion 
rates reproduce the qualitative shape of the distribution of gravitational 
torques. Observations of bar axis ratios give the same result. Thus, it seems
reasonable to conclude that galaxies accrete large amounts of gas, doubling 
their mass in less than one Hubble time, even if the exact value we have 
derived for the accretion rate may need fine-tuning in future studies.

\subsection{Other processes for bar evolution}

We have identified gas accretion as being the process 
responsible for bar reformation in spiral galaxies. 
Galaxy interactions may also explain the formation of bars or
strengthen existing bars (Gerin et al. 1990). 
The self-destruction of the first bar heats the disk up
and makes it stable. Galaxy interactions would not cool the disk down, a 
prerequisite for bar reformation and spiral arm maintenance. 
In the absence of accreted gas, the disk would then become stable because
of a large velocity dispersion, thus it would remain unbarred while the spiral 
structure would become weaker and weaker. Gas accretion, 
however, keeps the disk cold, and it is then much more susceptible to 
periodic bar reformation and spiral structure maintenance. 

Accretion is thus much more efficient compared to galaxy interactions in 
maintaining gravitational torques (spirals and bars) in disk galaxies. 
However, including the influence of galaxy interactions is in preparation. 
This will allow the distribution of bar strengths from our models (as well 
as the accretion rates we derive) to be refined, but interactions do not 
invalidate our present conclusions.

The case of lenticulars will be treated in a separate paper. 
S0s are stellar systems without much gas. 
In the absence of gas, the dynamics of disks is 
different: pure stellar bars are very robust, 
and can endure for one Hubble time (Combes \& Sanders
 1981), contrary to bars in spiral galaxies. In SOs, bars are not 
destroyed, and no mechanism is needed to explain bar reformation.

\section{Conclusion}\label{concl}
The issue of whether observations support a galaxy dissolving 
and reforming its bar has awaited the completion of
near-infrared surveys.  We are now able to show how robust this scenario is.
Isolated galaxies (non-accreting systems) cannot reproduce the observed
properties at all: they would become unbarred and spiral arms would 
disappear; many disks would then be nearly axisymmetric after a few Gyrs. 
On the contrary, spiral galaxies appear to be open
systems that are still forming and continuously accreting mass today.
We expect a doubling in disk mass every 10 billion years.
The origin of the accreted gas has not been considered, but the most
likely source could be the reservoirs of gas observed outside
 nearly all spiral disks (Sancisi 1983). 
Pfenniger et al. (1994) have even postulated 
that the dark matter around spiral galaxies might be in the form of 
cold gas. 
Accretion rates from infalling dwarf satellites only represent 
a few percent of the accretion rate that we derive (Toth \& Ostriker 1992),
so that other sources of accretion must be invoked.

\begin{acknowledgements}
We acknowledge the valuable comments from an anonymous referee. 
Funding for the OSU Bright Spiral Galaxy Survey was provided 
by grants from The National Science Foundation (grants AST-9217716 
and AST-9617006), with additional funding by the Ohio State University. 
We are indebted to P. Eskridge, M. Merrifield and L. Whyte for making 
the OSU sample available to us in digital format. 
We have made use of the LEDA database. 
The research of DLB is funded by the Anglo American Chairman's Fund and 
SASOL. RB is supported by NSF Grant AST-0205143 to the University of Alabama.
\end{acknowledgements}

\end{document}